Thoughts on new statistical procedures for age-period-cohort analyses[1]

Andrew Gelman, Department of Statistics and Department of Political Science, Columbia University

14 June 2008

1. Introduction: young Democrats and middle-aged Republicans

I first read about the age-period-cohort problem many years ago, but I didn't think seriously about it until recently, when I saw some survey results showing party identification by age (see Figure 1). Americans in their forties are the most Republican group, a pattern that I would attribute to this group coming of political age during the presidencies of Jimmy Carter and Ronald Reagan. Older Americans might be more likely to associate the Republicans with Richard Nixon and Watergate, whereas younger voters associate the Democrats with Bill Clinton and the Republicans with George W. Bush. Thus can the difference in popularity of successive presidents propagate into systematic patterns of political attitudes by age.

But do I really have the evidence to make this claim? Implicitly I am following a model in which voters lock in their party identification early, perhaps between the ages of 15 and 25, and then stay with this (perhaps with small changes) throughout their lives. How could we test this model? If we had longitudinal data, following up voters for decades, we could track the stability of their party identification over the years (and also try to identify the characteristics of those who switch and those who stay). Another approach would be to analyze repeated cross sections: If we were to make a graph, similar to Figure 1, but based on data from 2002, would the red and blue lines simply be shifted four years to the left, so that the party identification of 40-year-olds in 2006 matches that of the 36-year-olds in 2002? If such a pattern happened consistently over time, this would support the hypothesis of a *cohort effect*.

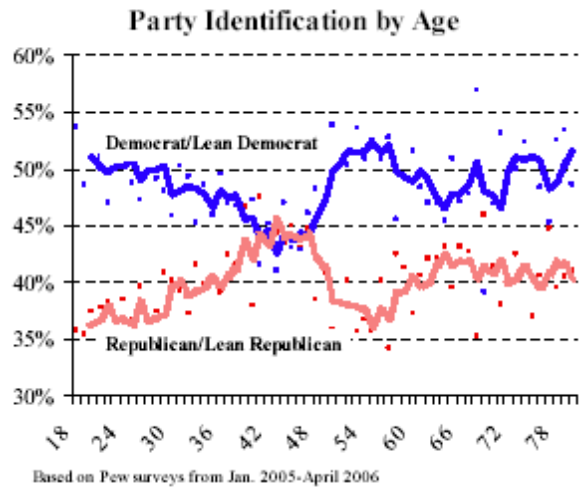

Figure 1. Survey data showing party identification in 2006 for different age groups, from Keeler (2006). The most Republican group is people in their mid-forties, who came of political age during Jimmy Carter's and Ronald Reagan's presidencies.

---

[1] Invited discussion for *American Journal of Sociology* of two papers on age-period-cohort analyses.

Another possibility that could be revealed by additional data is *an age effect*: Suppose that the equivalent to Figure 1, constructed four or eight or twelve years earlier, looked identical to the 2006 pattern with no age shift. Then we would be inclined to believe that party identification is associated with age, rather than cohort, with new voters starting out as strong Democrats, moving toward the Republican Party in their middle age, and then moving back to the Demcorats. It would not be difficult to construct a story consistent with this pattern (were it to in fact appear in the data), possibly associated with life-course changes involving marriage, children, and participation in the workforce.

Finally, the 2006 pattern may be part of a *period effect*. Figure 1 shows, overall, a strong Democratic advantage, but these polls were taken during a period when the Republicans have been on the defensive. Did the Democrats have such an advantage five or ten years ago? Maybe not. In this particular example, we are less interested in period effects—our primary goal here is to understand the big difference between today's young and middle-aged voters—but we certainly have to be aware of the possibility of period effects, if only to adjust for them in estimating age and cohort effects.

2. The difficulty of age-period-cohort analyses

Given this background, I was excited to have the opportunity to discuss this article on age-period-cohort analysis. I spoke with a colleague and formulated the following plan: I would send him a file with National Election Study data on party identification and age in each election year from 1948 through 2004, and he would then perform the following steps:
1. Create graphs of party identification vs. age for each year, following the pattern of Figure 1.
2. Fit the model described in the paper under discussion and see if the estimates make sense.
3. Play around with adding linear trends to the age, period, and cohort effects to get equivalent estimates that are equally consistent with the data.

The idea was to use the party identification problem as an applied test case for the new estimator. My further plan, after successful completion of the three steps above, was to review the literature on party identification by age and to understand our results in light of this literature and any available longitudinal surveys.

Unfortunately, for personal reasons my colleague was not able to carry out the above plan, nor did I have the time to do it myself. I have thus illustrated one of the key difficulties with age-period-cohort effects, which is the practical effort required to put together a dataset and fit a model. The tasks are not hugely complicated but neither are they as simple and codified as running a regression analysis. In this case, the bottom line is that my comments on the proposed new method will be restricted to the theoretical, and I will postpone my engagement with the age and voting literature to a later date (at which point, perhaps, young people will be voting Republican again).

3. Conclusions

The discussants of the paper at hand correctly note that there can be no general solution to the age-period-cohort problem: as has been noted by Fienberg and Mason (1979), Holford (1983), and many others, the likelihood function has a ridge that no amount of analytical manipulation can evade. For any dataset, there is a space of possible estimates, all equally consistent with the data, but with different linear trends in age, period, and cohort, and no way from the data alone to choose among these.

On the other hand, some of the estimates seem to make more sense than others. For example, consider a model which adds the following three trends: (i) since 1948, an increase of 1% per year in the overall probability of Democratic identification, (ii) starting at age 18, an increase of 1% per year in the probability of an individual being a Republican as he or she gets older, and (iii) for each cohort, a increase of 1% in the probability of being Republican, compared to the cohort that was born one year earlier. Add these three trends together and you get zero—the combination has no effect on any observable data—but they do not make much political sense. What does it really mean to talk about a linear time trend toward the Democrats if it is exactly canceled by each cohort being more Republican than the last?

To put it another way, some methods of constraining the possible space of solutions seem more reasonable than others. In the party identification example, it might make sense to assume a zero overall trend in period effects, or even a very slight trend toward the Republicans; this would make more sense than imposing a zero overall trend in age effects, or imposing some sort of arbitrary constraint on some endpoints, of the sort that is sometimes done to impose identifiability. For these reasons, I am open to persuasion that a method such as presented in the paper under discussion can be of practical use. My preference would be for something slightly more general—a framework for imposing sensible constraints, rather than a single automatic procedure—but I suspect that it is through examples that we will see if the method provides real benefits beyond existing approaches.

I conclude with a comment about possible Bayesian extensions. In classical likelihood inference, reparameterization has no effect and is merely a matter of convenience—in this case, determining a particular preferred solution to the model. In Bayesian inference, however, a change of parameters can change the model if the form of the prior distribution is fixed (Gelman, 2004). In practice, prior distributions and likelihoods are chosen for mathematical convenience. For example, it is standard to assume independence of parameters in the prior distribution, and thus a transformation that rotates parameter space can change posterior inferences. For that reason, I am not completely skeptical of the potential for new methods that can help us analyze age-period-cohort phenomena.

Acknowledgments

I thank Andy Abbott for inviting this article and the National Science Foundation, National Institutes of Health, and Columbia University Applied Statistics Center for financial support.

Additional references